\documentclass[12pt]{article}
\usepackage{epsfig}

\newcommand{\psibar} {\bar{\psi}}

\newcommand{\beqn} {\begin{equation}}
\newcommand{\eqn} {\end{equation}}

\newcommand{\scr}[1] {\mbox{\scriptsize #1}}
\newcommand{\mq} {m_{\scr{q}}}

\newcommand{\slsh}[1] {#1\kern-.43em/}

\newcommand{\real}{{\sf I}\kern-.12em{\sf R}}
\newcommand{\comp}{{\sf I}\kern-.48em{\sf C}}
\newcommand{\nin} {\in\kern-.6em/}

\newcommand{\tc} {T_{\scr{c}}}

\begin{document}
hep-lat/9503011 \hfill BI-TP 95/12 \\

\begin{center}
{\Large\bf DECONFINEMENT AND HOT HADRONS\\[1ex]
           IN CRAYS AND QUADRICS
\footnote{
Summary of invited talks presented at the Hirschegg Workshop XXIII
`Dynamical Properties
of Hadrons in Nuclear Matter', Jan. 16--21, Hirschegg, Austria and
`Chiral Dynamics in Hadrons and Nuclei', Feb. 6--10, Seoul, Korea}
\\[1ex]}
{\large G. BOYD \\[1ex]}
{Fakult\"at f\"ur Physik, Universit\"at Bielefeld \\
Postfach 100131, D-33501 Bielefeld, Germany}\\
\end{center}

\begin{abstract}
  The equation of state of pure QCD, obtained from lattice QCD, is discussed
  for temperatures ranging from $0.9\tc$ to $4\tc$, as well as results on
  screening masses, the chiral condensate, and the pion decay constant close to
  the deconfinement phase transition in the confined phase of QCD.  The
  equation of state differs significantly from that of a free gas.  There is
  little evidence of a temperature dependence in the chiral condensate or the
  meson properties, but perhaps some for the nucleon screening mass. Above the
  phase transition one sees non-perturbative effects, even though hadron
  correlators show the existence of deconfined quarks.
\end{abstract}

\section{Introduction}

The study of hadronic matter at high temperature and / or density is receiving
increasing attention, both experimentally and theoretically.
It is important to gain a good understanding of the predictions of QCD in
this regime as soon as possible.
There are many approaches to this task, some of which are presented
elsewhere in these
proceedings. Each is advantageous for some aspect
of the problem.
One needs to combine them all to understand what QCD
predicts over the entire range of temperatures and densities relevant for heavy
ion collisions.

Recent results from one of the major theoretical players, the lattice
regulation of QCD at finite temperature,
will be presented below. An introduction to lattice QCD may
be found, for example, in~\cite{MONTMUEN,FKHWA}.

When one is close to the phase transition, many of the traditional approaches
to QCD run into problems, as both low temperature and high temperature
expansions need to be extrapolated into regions far from the regions in which
they are known to be secure. The lattice may well be the only satisfactory
approach in this region.

The equation of state, both for nuclear matter just below the deconfinement
transition, and for the quark-gluon plasma above the phase transition, is
required information for, amongst others, hydrodynamic models attempting to
describe the evolution of the quark gluon plasma.
Since the couplings and masses are not small below $2\tc$
the lattice approach is the obvious method to use.

Despite extensive studies of the QCD phase
transition~\cite{FKHWA,BP92KRH93}
%~\cite{DTK87,GLT87,GRH88,BGI91,STAM93,GBWALL}
little is known with certainty about the changes
in the excitation spectrum. In particular,
the properties of the (quasi)-particle spectrum in the confined phase
close to $\tc$ require a more thorough
understanding, as
the temperature dependence of hadron masses and other hadronic parameters
will lead to observable consequences in current, and especially forthcoming,
heavy ion collisions.

The temperature dependence of various hadronic properties has been addressed
using a number of different approaches.  These calculations yield different
predictions for the behaviour of some
quantities~\cite{LEUT90AND}.  Since these
approaches have limited applicability at high temperatures, one may hope that a
lattice calculation of the variation with temperature of any of these
quantities below the phase transition may shed  light on the discrepancies.

Another important quantity is the ratio of the chromo electric to chromo
magnetic screening length. This is used, for example, in estimates of the
colour transparency of the plasma~\cite{GYU93}. One promising approach is to
calculate it directly from measurements of the gluon self energy on the
lattice.

Recent results on the equation of state for pure gluon SU(3) theory and for QCD
with two light flavours will be presented in section~\ref{sec:eos}.  The
results from a study of hadronic properties in quenched QCD at temperatures
between $0.75\tc$ and $0.92\tc$ are presented in section~\ref{sec:hadrons}.
Section~\ref{sec:gluon} contains the results pertaining to the chromo electric
and magnetic screening masses.

\section{Equation of State}
\label{sec:eos}

Recent work~\cite{ENGELS90EKR94} has enabled the calculation of the energy
density and pressure on the lattice using an entirely non-perturbative
technique for the first time. This requires detailed a knowledge of
$dg^{-2}/da$, ie. one needs to know the QCD
beta function $\beta_{\scr{f}}$ relating changes in physical quantities
with changes in the coupling. This has been calculated by the TARO
collaboration~\cite{TARO93}.

\begin{figure}[tb]
   \begin{minipage}{0.48\textwidth}
  \begin{center}
    \leavevmode
    \epsfig{bbllx=210,bblly=300,bburx=500,bbury=500,
        file=e3p.ps}
   \caption{The interaction measure $(\epsilon - 3P)/T^{4}$  against
            temperature for three different values of the temporal
            extent $N_\tau$ in pure SU(3) gauge theory.
           }
   \label{fig:interaction}
   \end{center}
   \end{minipage}
   \hfill
   \begin{minipage}{0.48\textwidth}
  \begin{center}
    \leavevmode
    \epsfig{bbllx=210,bblly=300,bburx=500,bbury=500,
        file=efk.ps}
   \caption{The energy density compared to the energy density in the
            Stefan-Boltzmann limit on a lattice of the same size against the
            temperature in units of the critical temperature.
           }
   \label{fig:evsesb}
   \end{center}
   \end{minipage}
\end{figure}

Given the beta function one calculates the interaction measure,
the difference between the energy density and thrice the pressure,
as follows:
\begin{equation}
\label{eq.ta}
\frac{\epsilon-3p}{T^4}  =  T \frac{\partial}{\partial T}
                              \left( \frac{p}{T^4} \right)
        =  a \left( \frac{d g^{-2}}{d a} \right)
             6N_c N_{\tau}^4 \{ P_{\sigma} + P_{\tau} - 2 P_0 \}.
\end{equation}
The pressure is given by
\begin{equation}
\frac{p}{T^4} =
\left. \frac{-f}{T^4} \right|^{\beta}_{\beta_0}
       = 3 N_{\tau}^4\int_{\beta_0}^{\beta} d\tilde \beta \,
             \{ P_{\sigma}(\tilde \beta) +
                P_{\tau}(\tilde \beta) - 2 P_0(\tilde \beta) \}
\label{eq.fint}
\end{equation}
The zero temperature plaquette is given by $P_0$, the spatial and temporal
plaquettes at finite temperature are given by $P_{\sigma}$ and $P_{\tau}$
respectively.  Both the energy density and the pressure are normalised to zero
at zero temperature.

The results obtained for the interaction measure, divided by $T^{4}$,
are shown in figure~\ref{fig:interaction}. The peak in the
interaction measure shortly after the first order deconfinement phase
transition shows clearly the strong deviation from ideal gas behaviour,
$\epsilon = 3P$. The energy density rises sharply, whilst the
pressure lags behind until one has reached temperatures around $3\tc$. This
is evident again in figure~\ref{fig:ep} where the energy density and pressure,
scaled with $T^{4}$ are plotted together.

Note that the interaction measure is related to the value of the gluon
condensate at finite temperature~\cite{MILLER94,LEUT92} via
\begin{equation}
\langle G^{2} \rangle_{T} = \langle 0 | G^{2} | 0 \rangle
                           - (\epsilon - 3P)_{T},
\end{equation}
where $\langle G^{2} \rangle_{T}$ and $\langle 0 | G^{2} | 0 \rangle$ are the
condensates at zero and finite temperature respectively. The gluon field
strength $G^2$ is given by $G^{2} =
-[\beta_{\scr{f}}/(2g^{3})]G^{\mu\nu}_{a}G^{a}_{\mu\nu}$.  Taking $\langle 0 |
G^{2} | 0 \rangle=0.0135$GeV$^{4}$~\cite{LEUT92} one finds that
$\langle G^{2} \rangle_{T}$
drops to zero immediately after the phase transition, and then goes negative
according to the above equation.

The energy density for different values of the temporal extent is plotted in
figure~\ref{fig:evsesb}, compared to the Stefan-Boltzmann limit on this size
lattice. If one takes the continuum limit one obtains a value around
83\% of the Stefan-Boltzmann value for temperatures above $2\tc$ up to at least
$3\tc$. It is clear that even at these temperatures the interactions and masses
of the excitations cannot be neglected.  A similar rapid rise in the energy
density, and slower rise in the pressure, is seen in the case of two flavour
QCD~\cite{BKT94}.

\begin{figure}[t]
   \begin{minipage}{0.49\textwidth}
  \begin{center}
    \leavevmode
    \epsfig{bbllx=210,bblly=300,bburx=500,bbury=500,
      file=e3pt4.ps}
   \caption{The energy density and pressure in units of $T^4$ on
            a lattice of size $32^{3} \times 6$. Also shown is the
            continuum Stefan-Boltzmann limit (lower solid line)
            and the Stefan-Boltzmann limit for this size lattice
            (upper solid line).
           }
   \label{fig:ep}
   \end{center}
   \end{minipage}
   \hfill
   \begin{minipage}{0.49\textwidth}
  \begin{center}
    \leavevmode
    \epsfig{bbllx=205,bblly=260,bburx=470,bbury=580,
        file=vos.ps,height=74mm,angle=-90}
   \caption{The speed of sound squared, $c^{2}_{\scr{s}}=dP/d\epsilon$ as a
            function of temperature on lattices of size $16^{3}\times 4$
            and $32^{3}\times 6$ as a function of temperature.
           }
   \label{fig:sound}
   \end{center}
   \end{minipage}
\end{figure}

It is clear that, if there is a jump in the energy density, the pressure will
lag behind in a thermodynamic system, from the relation $P/T = P_{0}/T + \int
dT \epsilon / T^{2}$. The deviation from the ideal gas behaviour may be
understood in terms of plasma consitituents that are massive, and have a strong
interaction~\cite{RIS92}.

The speed of sound in a medium is another important parameter. This is shown
in figure~\ref{fig:sound} for the high temperature phase of pure gauge theory.
The speed of sound also approaches the ideal gas value ($1/\sqrt{3}$),
but is only at around 80\% of it even at $2\tc$.

\section{Hadronic Properties}
\label{sec:hadrons}
\subsection{Chiral Sector}
\label{sec:chiral}
The quark condensate, the quantity from which other approaches obtain much of
the effect of temperature on the hadrons, is the logical place to start.  It is
known to change drastically at the deconfinement transition in quenched
QCD.

\begin{figure}[tbhp]
   \begin{minipage}{75mm}
  \begin{center}
    \leavevmode
    \epsfig{bbllx=100,bblly=250,bburx=500,bbury=650,
        file=psi_m_60.ps,height=70mm,angle=-90}
   \end{center}
   \end{minipage}
\hfill
\begin{minipage}{75mm}
  \begin{center}
    \leavevmode
      \epsfig{bbllx=100,bblly=250,bburx=500,bbury=650,
          file=psi.ps,height=70mm,angle=-90}
     \end{center}
   \end{minipage}
         \caption{   \label{fig:ccvsmq}      \label{fig:chcond}
           Figure (a) shows the chiral condensate as a function of the bare
           quark mass, at $\beta = 6.00$.  The upper curve is obtained using
           the value taken directly from the trace of the fermion matrix, the
           lower curve after subtracting the linear dependence on the quark
           mass.  The results have been taken
           from~\protect\cite{SUBTC94,GGKS91}.  Figure~(b) shows the finite
           temperature chiral condensate normalised to its value at zero
           temperature.  The circles represent quenched data, extrapolated
           to zero quark mass
           from~\protect\cite{SUBTC94}. The other figures represent results
           for various numbers of flavours, but not extrapolated to zero
           quark mass. }
\end{figure}

The chiral condensate is shown in figure~\ref{fig:ccvsmq}(a) as a function of
bare quark mass at $T=0.92\tc$. It may be calculated directly from the trace of
the fermion matrix ($\langle\psibar\psi\rangle_{\rm SE}$) and by using the pion
and sigma propagators, which corrects for the linear dependence on the quark
mass~\cite{KPS87}. One can see quite clearly that both methods extrapolate to
the same value at zero quark mass. The same holds true for lower temperatures.

The temperature dependence of the chiral condensate is shown in
figure~\ref{fig:chcond}(b), where the ratio of the finite temperature chiral
condensate to that at $T=0$ is plotted. These values have been extrapolated to
zero quark mass, which is the reason why the error bars are so large. If one
does not extrapolate, the ratio remains the same, but with errors a factor of
ten smaller.  For comparison results for full QCD,
including dynamical fermions, have been included.  Since none of these results
are at the physical quark mass, the effect of temperature is probably
underestimated.  It is clear that the chiral condensate does not change
significantly until one is extremely close to the critical temperature.

The pion mass displays the behaviour expected of a
Goldstone particle, $m_{\pi}^{2} = A_{\pi}\mq$, at all temperatures. There is
also no sign of a dependence on temperature in the pion mass, or in the
slope $A_{\pi}$.

Another quantity of considerable interest for understanding the temperature
dependence of the chiral sector of QCD is the pion decay constant.
It can be determined directly from the relevant matrix element on the
lattice~\cite{KPS87}.
At zero temperature $f_\pi$ is related to the
chiral condensate and the pion mass
through the Gell-Mann--Oakes--Renner relation,
which is expected to be valid as long as chiral
symmetry remains spontaneously broken.
\begin{equation}
m_\pi^2 f_\pi^2\;=\;\mq\langle \bar{\psi}\psi \rangle_{(\mq=0)} ~~.
\label{eq:gmor}
\end{equation}

The temperature independence of the amplitude, $A_\pi$, and the chiral
condensate up to $T=0.92\tc$ indicates that the pion decay constant will also
be temperature independent.  A direct calculation shows that it is, and that
the GMOR relation holds, for temperatures up to at least $T=0.92\tc$.

\subsection{Nucleon and meson masses}
\label{sec:nucrho}
One obtains hadron screening masses by fitting the exponential decrease of the
correlator $C_{\scr{H}}(z)$ to, for example, an hyperbolic cosine. One obtains
either local masses, if only two points are fitted to, or an estimate of the
lowest state if one fits to the long distance part of the correlator. The
lowest mass which can be
extracted from the fit, $E_{\scr{H}}$, is related to the screening mass for
fermionic states via
\begin{equation}
E_{\scr{H}}^{2} = m_{\scr{H}}^{2}
                + k \sin^{2}\left( \pi/N_{\tau} \right) ,
\label{eq:mn}
\end{equation}
due to the contribution from the non-vanishing Matsubara energy, $p_0 = \pi T$.
Note that for bosons $E_{\scr{H}}^{2} = m_{\scr{H}}^{2}$.

The local rho screening masses and the value obtained from a fit to the full
propagator at $T=0.92\tc$ and at zero temperature are shown in
figure~\ref{fig:mrholoc}(a).  It is clear that the rho screening mass does not
show any significant dependence on temperature.

The nucleon has been examined using wall sources at a quark mass of 0.01 in
lattice units.  The local screening masses for $T=0.92\tc$, and the screening
mass obtained from a full fit, are shown in figure~\ref{fig:mnloc}(b) along
with
the result obtained at zero temperature.  The increase in temperature clearly
has a dramatic effect on the nucleon.  However, most of this can be understood
purely in terms of the fact that the lowest momentum of the nucleon is not
zero, but $\pi T$, as discussed above.

The screening mass at $T=0.92\tc$, extracted using eqn.~\ref{eq:mn} with the
assumption that $k=1$, indicates that the mass rises slightly:
$m_{\scr{n}}(0.92\tc) = (1.1\pm 0.03)m_{\scr{n}}(T=0)$. We note, however, that
there is also the possibility of a modification of the energy dispersion
relation at finite temperature, which may lead to a deviation of $k$ from unity
\cite{LES90}.  The error given above is from the statistical error alone, as we
cannot determine the systematic error which may be introduced by our assuming
that eq.~\ref{eq:mn} with $k = 1$ is applicable.

\begin{figure}[hbtp]
  \begin{minipage}{75mm}
  \begin{center}
    \leavevmode
    \epsfig{bbllx=100,bblly=250,bburx=500,bbury=650,
        file=lmrho600.ps,height=70mm,angle=-90}
  \end{center}
  \end{minipage}
  \hfill
  \begin{minipage}{75mm}
  \begin{center}
    \leavevmode
    \epsfig{bbllx=100,bblly=250,bburx=500,bbury=650,
        file=lmn600.ps,height=70mm,angle=-90}
  \end{center}
  \end{minipage}
  \caption{   \label{fig:mrholoc} \label{fig:mnloc}
        Figure (a) shows
        the local rho screening mass at $T=0.92\tc$, and
           $\mq = 0.01$ in lattice units, from wall sources. The solid and
           dotted lines show the fitted value and errors, the long dashed
           lines the zero temperature values from~\protect\cite{KIM93}.
        Figure (b) shows
        the local nucleon screening mass at $T=0.92\tc$, and
           $\mq = 0.01$ in lattice units, from wall sources. The solid and
           dotted lines show the fitted value and errors, the long dashed
           lines the zero temperature values from~\protect\cite{KIM93}.
          }
\end{figure}

Finally, one would like to know the properties of correlators with hadronic
quantum numbers in the deconfined phase. These can be studied using correlators
in the spatial direction, as above. One finds that the nucleon then approches
thrice the lowest Matsubara frequency, and all mesons other than the pion
approach twice the lowest Mastubara frequency (see~\cite{MTC90} and refernces
therein). This is what one expects if the quarks become deconfined.

The pion, though, acquires a screening mass half as large as expected for a
free meson. As it is the particle associated with chiral symmetry, one may
expect it to have a special role even above the phase transition~\cite{HAK85}.
In order to understand whether the pion is indeed deconfined as well, with a
different interaction between the two quarks, the meson correlators were
examined using different techniques~\cite{SPAT94}. In one of these, temporal
correlators generated with wall sources were examined. Temporal correlators
show the true mass, and not the screening mass. Wall sources are used in order
to project out only the contribution from the lowest state. The results for the
local masses are shown in Figure~\ref{fig:tempwall}, along with results for the
effective quark mass.

\begin{figure}[tbp]\leavevmode
\centerline{\epsfbox[0 0 387 240]{tempwall.ps}}
\caption{Local masses $m_{\scr{H}}^{\scr{W (=wall)}}$ from correlators
  constructed with wall sources in the PS (circles) and V (squares) channels in
  the high temperature phase.  Also shown is $2m_q^{\scr{eff}}$ (triangles)
  from \protect\cite{BGK92}.  Lines have been drawn to guide the eye.  }
\label{fig:tempwall}
\end{figure}

One sees here that both the rho and the pion masses tend towards twice the
effective quark mass at high temperatures.
This, combined with other investigations on the lattice, indicate that the
hadrons (including the pions) become deconfined above $\tc$, with rather large
channel dependent residual interactions.

One does, in all cases, see the factor
of two difference between the pion and rho masses or screening masses.
There is clearly a difference between a quark--anti-quark pair carrying pion
quantum numbers and one carrying rho quantum numbers. Explanations for the
relative lightness of the pion have been
put forward based on spin--spin interactions~\cite{KOCH92} and residual gluon
condensates~\cite{IHAT94}.

\section{Electric and Magnetic Screening Masses}
\label{sec:gluon}

The chromo electric screening mass can be obtained by examining, amongst
others, the correlation between Polyakov loops or the pole of the gluon
correlator.  At high temperature in perturbation theory at next to leading
order it has been shown that the two approaches yield the same
result~\cite{REB94,BRN94}.

The chromo magnetic mass cannot be determined in perturbation theory, due to
infra-red divergences. Preliminary results from calculations in
progress~\cite{BIELWIP}, in pure gauge theory, indicate that the magnetic and
electric screening masses have the same order of magnitude above the phase
transition, and for temperatures as high as $2\tc$. This result is consistent
with the results obtained in another study in SU(2) pure gauge
theory~\cite{RANK}.

This can be compared with the results from the potential, and the spatial
string tension~\cite{ED94} which indicate that the coupling remains large well
above the phase transition, with $g(T=2\tc)\approx 2$.

\section{Conclusions}
\label{sec:conc}

The chiral condensate does not show any significant temperature dependence up
to $T= 0.92\tc$. In view of this the observation that there is no sign of a
temperature dependence in the pion decay constant, pion screening mass, sigma
screening mass or the rho screening mass may not come as a surprise.  There is
some, albeit inconclusive, indication of a temperature dependence for the
screening mass of the nucleon.

The details of the temperature dependence may change in the case of QCD with
two flavours, where the transition is expected to be second order rather than
first order as in the the pure $SU(3)$ gauge theory, and with quark masses
close to the physical quark mass. However, these differences will probably not
show up for temperatures less than $0.9\tc$.

The equation of state for pure gauge theory is now reasonably well understood.
One finds clear evidence for a departure from the ideal gas at below $2\tc$, at
which point the energy density reaches only 83\% of the Stefan-Boltzmann limit.
Thus it is clear that there are strong interactions, or equivalently a large
contribution from non-perturbative effects, above the phase transition.
This is supported by measurements of the spatial string tension, which indicate
a coupling $g\approx 2$, results for the chromo electric and magnetic
screening masses, as well as the effective quark mass in the Landau gauge.

All states with hadronic quantum numbers appear to be deconfined above the
phase transition. However, the coupling between the quark and anti-quark
depends on the channel considered, leading to pion / sigma masses of around
half the rho mass.

\section*{Acknowledgements}

The results discussed here have been supported in part by the Stabsabteilung
Internationale Beziehungen,Kernforschungszentrum Karlsruhe, a NATO research
grant, contract number CRG 940451 and a DFG grant, DFG Pe-340/1, DFG
Pe-340/6-1,
and the HLRZ in J\"ulich. I would like to thank the organisers of the
conference for the invitation and the support received.  I would also like to
thank my colleagues, J. Engels, S. Gupta, F. Karsch, E. Laermann, C. Legeland,
M. L\"utgemeier, B.  Petersson and K. Redlich for enlightening discussions and
productive collaboration.

\end{document}